\documentclass[twocolumn,amsmath,amssymb,superscriptaddress,notitlepage]{revtex4-1}
\usepackage{graphics}
\usepackage{bm}
\usepackage{dcolumn}
\usepackage{natbib}
\usepackage{epstopdf}
\usepackage{float}
\usepackage{graphicx}
\usepackage{epsfig}
\usepackage[pdfstartview=FitH]{hyperref}
\usepackage{color}
\usepackage{appendix}
\usepackage{ulem}

\newcommand{\rr}{{\bf r}}

\begin{document}
\title{Halperin $(m',m,n)$ fractional quantum Hall effect in topological flat bands}
\author{Tian-Sheng Zeng}
\affiliation{Department of Physics, College of Physical Science and Technology, Xiamen University, Xiamen 361005, China}
\date{\today}
\begin{abstract}
The Halperin $(m',m,n)$ fractional quantum Hall effects of two-component quantum particles are studied in topological checkerboard lattice models. Here for $m\neq m'$, we demonstrate the emergence of fractional quantum hall effects with the associated $\mathbf{K}=\begin{pmatrix}
m+2 & 1\\
1 & m\\
\end{pmatrix}$ matrix (even $m=2$ for boson and odd $m=3$ for fermion) in the presence of both strong intercomponent and intracomponent repulsions.
Through exact diagonalization and density-matrix renormalization group calculations, we elucidate their topological fractionalizations, including (i) the $\det|\mathbf{K}|=(m^2+2m-1)$-fold ground-state degeneracies and (ii) fractionally quantized topological Chern number matrix $\mathbf{C}=\mathbf{K}^{-1}$. Our flat band model provides a paradigmatic example of a microscopic Hamiltonian featuring fractional quantum Hall effect with partial spin-polarization.
\end{abstract}
\maketitle

\section{Introduction}

Multicomponent fractional quantum Hall (FQH) effect offers us a new scope for realizing new kind of topological quantum phases, which expands the taxonomy of topological phases of matter~\cite{Wen2017}. Here the term ``multicomponent" refers to particle systems with an internal degree of freedom, like spin, layer, valley and so on. The simplest example of a multicomponent quantum Hall system would be Halperin's two-component quantum Hall effect~\cite{Halperin1983} and quantum Hall ferromagnet~\cite{Wen1992}. Nevertheless they may host nontrivial topological excitations like skyrmion~\cite{Moon1995,Yang1996} that have no analogue in one-component systems. Besides, early experimental observation of $\nu=\nu_{\uparrow}+\nu_{\downarrow}=1/2$ FQH effect in bilayer systems~\cite{Suen1992}, has been proposed to be the likely Halperin (331) FQH state~\cite{Chakraborty1987,Yoshioka1988,Yoshioka1989,He1991,He1993}. Subsequently, when valley degrees of freedom are included, for instance in graphene sheets with the fourfold spin-valley degeneracy, a rich class of four-component FQH effects is experimentally discovered~\cite{Bolotin2009,Dean2011}, serving as an effective SU(4) generalization of Halperin's FQH effect~\cite{Toke2007,Goerbig2007,Gail2008}. In two parallel graphene layers separated by a thin insulator, the layer degree of freedom endows us several new types of interlayer-correlated FQH effects~\cite{Liu2019,Li2019}.
Meanwhile, the mimicking of Landau level in rapidly rotating Bose gases also inspired the search for bosonic FQH effects~\cite{Cooper2020}. With pseudospin degree of freedom, a lot of multicomponent bosonic quantum Hall effects like Halperin $(221)$ FQH effect~\cite{Grass2012,Furukawa2012,Wu2013,Grass2014} and bosonic integer quantum Hall effect~\cite{Senthil2013,Furukawa2013,Regnault2013,Wu2013,Grass2014}, have been actively explored.

Further, in the past decade, the arise of topological flatbands in analogy to two-dimensional Landau levels, has became an exciting platform for studying the quantum Hall effect in two-dimensional lattice systems with broken time-reversal symmetry (now called Chern insulators)~\cite{Liu2023}. In particular, single-component FQH effects at specific fillings  $\nu=1/(C+1)$ (for hardcore bosons) and $\nu=1/(2C+1)$ (for spinless fermions) in topological flat bands with higher Chern numbers $C>1$, are believed to be color-entangled lattice versions of multicomponent FQH effect~\cite{LBFL2012,Wang2012r,Yang2012,Sterdyniak2013,YLWu2013,YLWu2014,YHWu2015,Andrews2018,Andrews2021}. For multicomponent quantum particles in topological flatbands with Chern number $C=1$, strong intercomponent and intracomponent two-body repulsions lead to the emergence of many multicomponent Abelian FQH effects~\cite{Zeng2017,Zeng2018,Zeng2019,Zeng2020,Zeng2022a} (including two-component FQH effects of Bose-Fermi mixtures where quantum particle statistics is tacitly regarded as an internal degree of freedom~\cite{Zeng2021}) and bosonic integer quantum Hall effect~\cite{Zeng2022}. Extensive numerical calculations of the Chern number matrices of these degenerate ground states reveal their $\mathbf{K}$-matrix classifications within the framework of the Chern-Simons gauge-field theory~\cite{Wen1992a,Wen1992b,Blok1990a,Blok1990b,Blok1991}, where one can derive the $\mathbf{K}$ matrix from the inverse of Chern number matrix. Further, it is numerically proposed~\cite{Zeng2022b} that two-component bosons in topological $C=1$ flatbands may host a non-Abelian spin-singlet FQH effect at filling $\nu=\nu_{\uparrow}+\nu_{\downarrow}=4/3$ in the presence of strong intercomponent and intracomponent three-body repulsions~\cite{Ardonne1999}. Together with Landau levels, topological flat bands offer us an alternative opportunity of realizing multicomponent quantum Hall systems, with many possible FQH states to be explored.

In this paper, we investigate the emergence of Halperin $(m',m,n)$ FQH effects~\cite{Seidel2008} of two-component quantum particles in a microscopic topological lattice model with the lowest band carrying unit Chern number $C=1$. We focus on the numerical identification of the $\mathbf{K}$ matrix with unequal matrix elements $m\neq m'$ based on Chern-number matrix structure and nontrivial topological degeneracy.

This paper is organized as follows. In Sec.~\ref{model}, we introduce the microscopic interacting Hamiltonian of two-component quantum particles loaded on topological $\pi$-flux checkerboard lattice, and give a description of our numerical methods. In Sec.~\ref{Halperin}, we study the many-body ground states of these two-component particles in the presence of strong intercomponent and intracomponent two-body repulsions, present numerical results of the $\mathbf{K}$ matrix by exact diagonalization at fillings $\nu_{\uparrow}=1/7,\nu_{\downarrow}=3/7$ for spinful hardcore bosons and $\nu_{\uparrow}=1/7,\nu_{\downarrow}=2/7$ for spinful fermions. In Sec.~\ref{dmrg}, we complement our topological characterization by calculating the fractional charge pumping, entanglement spectrum of the ground states from density-matrix renormalization group. Finally, in Sec.~\ref{summary}, we summarize our results and discuss the prospect of investigating nontrivial topological states in multicomponent systems.

\section{Model and Method}\label{model}

\begin{figure}[b]
  \includegraphics[height=2.6in,width=3.0in]{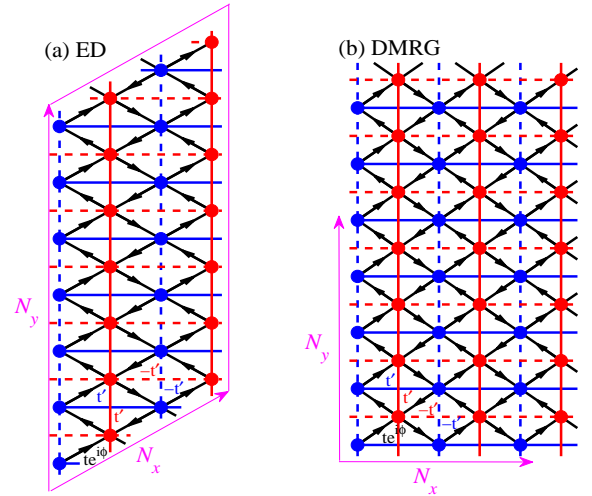}
  \caption{\label{lattice}(Color online) The schematic plot of topological $\pi$-flux checkerboard lattice used by ED and DMRG studies. The magenta arrows indicate the real-space lattice translational direction. The black arrow direction present the signs of the phases $\phi$ in the nearest-neighbor hopping. The blue (red) filled circles represent the sublattice sites. The arrow link shows the hopping direction carrying chiral flux $\phi_{\rr'\rr}$. The next-nearest-neighbor hopping amplitudes are $t_{\rr,\rr'}'=\pm t'$ along the solid (dotted) lines.
  }
\end{figure}

Here, we will numerically demonstrate the emergence of partially spin-polarized FQH effect of two-component quantum particles (either hardcore bosons or fermions) with strong interactions in topological flat bands through density-matrix renormalization group (DMRG) and exact diagonalization (ED) simulations. Two-component quantum particles are coupled with each other via extended Hubbard interaction in prototypical topological $\pi$-flux checkerboard (CB) lattice~\cite{Sun2011} whose lowest band host a Chern number $C=1$. The model Hamiltonian is given by
\begin{align}
  H&=\!\sum_{\sigma}\!\Big[-t\!\!\sum_{\langle\rr,\rr'\rangle}\!e^{i\phi_{\rr'\rr}}c_{\rr',\sigma}^{\dag}c_{\rr,\sigma}
  -\!\!\!\!\sum_{\langle\langle\rr,\rr'\rangle\rangle}\!\!\!t_{\rr,\rr'}'c_{\rr',\sigma}^{\dag}c_{\rr,\sigma}\nonumber\\
  &-t''\!\!\!\sum_{\langle\langle\langle\rr,\rr'\rangle\rangle\rangle}\!\!\!\! c_{\rr',\sigma}^{\dag}c_{\rr,\sigma}+H.c.\Big]+U\sum_{\rr}n_{\rr,\uparrow}n_{\rr,\downarrow}\nonumber\\
  +&\sum_{\sigma}\!\sum_{\langle\rr,\rr'\rangle}V_{\sigma\sigma}n_{\rr',\sigma}n_{\rr,\sigma}
  +\sum_{\sigma}\!\sum_{\langle\langle\rr,\rr'\rangle\rangle}\!\!V_{\sigma\sigma}'n_{\rr',\sigma}n_{\rr,\sigma},\label{cbl}
\end{align}
where $\langle\ldots\rangle$,$\langle\langle\ldots\rangle\rangle$ and $\langle\langle\langle\ldots\rangle\rangle\rangle$ denote the nearest-neighbor, next-nearest-neighbor, and next-next-nearest-neighbor pairs of sites, with the corresponding tunnel couplings $t'=0.3t,t''=-0.2t,\phi=\pi/4$ respectively (see the Figure~\ref{lattice} for detailed lattice geometry). Here $\sigma=\uparrow,\downarrow$ denote the spin indices for hardcore bosons (fermions), $c_{\rr,\sigma}^{\dag}$ is the particle creation operator at site $\rr=(x,y)$, $n_{\rr,\sigma}=c_{\rr,\sigma}^{\dag}c_{\rr,\sigma}$ is the particle number at site $\rr$. $U$ is the strength of the onsite Hubbard repulsion, $V_{\sigma\sigma}$ is the strength of the nearest-neighbor repulsion, and $V_{\sigma\sigma}'$ is the strength of the next-nearest-neighbor repulsion.

Provided the Abelian two-component topological phase, it should be classified by the $\mathbf{K}=\begin{pmatrix}
m' & l\\
l & m\\
\end{pmatrix}$ matrix
with the variational correlation structure in two-dimensional complex coordinate space $z_j^{\sigma}=x_j^{\sigma}+iy_j^{\sigma}$ of the $j$-th particles with spin-$\sigma$ ($j=1,2,\cdots,N_{\sigma}$, and suppose the positive values $m',m,l>0$)
\begin{align}
  \psi\propto\prod_{i<j}(z_i^{\uparrow}-z_j^{\uparrow})^{m'}(z_i^{\downarrow}-z_j^{\downarrow})^{m}\prod_{i,j}(z_i^{\uparrow}-z_j^{\downarrow})^{l}\nonumber
\end{align}
where the power law exponents of intracomponent correlations $m,m'$ are even for bosons and odd for fermions due to diverse quantum particle statistics, similar to those in Laughlin states, while the exponent $l$ describes the intercomponent correlation, originating from the intercomponent interactions. Compared to the single-particle wave function $\psi_s\sim z^m$ with an orbital angular momentum $m$ (in units of $\hbar$) in the lowest Landau level under symmetric gauge, we obtain that the first spin-$\uparrow$ particle $z_1^{\uparrow}$ hosts the highest orbital angular momentum $m'(N_{\uparrow}-1)+lN_{\downarrow}$ and the first spin-$\downarrow$ particle $z_1^{\downarrow}$ also hosts the highest orbital angular momentum $m(N_{\downarrow}-1)+lN_{\uparrow}$. For a given number of orbital states $N_{\phi}$ and large particle numbers $N_{\uparrow},N_{\downarrow}\gg1$, $N_{\phi}\simeq m'N_{\uparrow}+lN_{\downarrow}\simeq mN_{\downarrow}+lN_{\uparrow}$, and we can derive the spin-up filling $\nu_{\uparrow}=N_{\uparrow}/N_{\phi}=(m-l)/(mm'-l^2)$ and spin-down filling $\nu_{\downarrow}=N_{\downarrow}/N_{\phi}=(m'-l)/(mm'-l^2)$. Therefore in lattice models, we fix the particle fillings of the lowest Chern band as $\nu_{\uparrow}=2N_{\uparrow}/N_s,\nu_{\downarrow}=2N_{\downarrow}/N_s$, where $N_{\sigma}$ is the global spin-$\sigma$ particle number with $U(1)$-symmetry, and $N_s$ is the total number of lattice sites. In the ED study of small finite periodic lattice systems with $N_x\times N_y$ unit cells ($N_s=2\times N_x\times N_y$), the energy states are labeled by the total momentum $K=(K_x,K_y)$ in units of $(2\pi/N_x,2\pi/N_y)$ in the Brillouin zone. For larger systems, we exploit infinite DMRG on the cylindrical geometry with finite width $N_y$ and infinite length $N_x\rightarrow\infty$, keeping the maximal bond dimension up to $M=10000$. In the DMRG, the geometry of cylinders is open boundary condition in the $x$ direction and periodic boundary condition in the $y$ direction.

\section{Halperin's Fractional Quantum Hall Effect}\label{Halperin}

In this section, we first systematically present numerical results for the topological information of the many-body ground state of the model Hamiltonian Eq.~\ref{cbl}. We show that (i) two-component bosonic FQH effect emerges at fillings $\nu_{\uparrow}=1/7,\nu_{\downarrow}=3/7$ under strong interactions $U,V_{\uparrow\uparrow}\gg t,V_{\downarrow\downarrow}=V_{\sigma\sigma}'=0$, whose topological order is characterized by the $\mathbf{K}_b=\begin{pmatrix}
4 & 1\\
1 & 2\\
\end{pmatrix}$ matrix, and (ii) two-component fermionic FQH effect emerges at fillings $\nu_{\uparrow}=1/7,\nu_{\downarrow}=2/7$ under strong interactions $U,V_{\sigma\sigma},V_{\uparrow\uparrow}'\gg t,V_{\downarrow\downarrow}'=0$, whose topological order is characterized by the $\mathbf{K}_f=\begin{pmatrix}
5 & 1\\
1 & 3\\
\end{pmatrix}$ matrix.

In topological checkerboard lattice model, extensive numerical calculations of the emerging FQH effect show that (i) for single component hardcore bosons, the addition of intracomponent nearest-neighbor repulsion $V$ enhances the stability of the Laughlin state with intracomponent particle correlation $\psi\propto(z_i-z_j)^4$~\cite{Wang2011}; (ii) for spinless fermions, the intracomponent nearest-neighbor repulsion $V$ only induces the Laughlin state with intracomponent particle correlation $\psi\propto(z_i-z_j)^3$ while the addition of the intracomponent next-nearest-neighbor repulsion $V'$ stabilizes the Laughlin state with intracomponent particle correlation $\psi\propto(z_i-z_j)^5$~\cite{Sheng2011}. Therefore, following the previous merit, when spin degree of freedom is included, we utilize asymmetric intracomponent interaction in Eq.~\ref{cbl} ($V_{\uparrow\uparrow}\gg V_{\downarrow\downarrow}$ for hardcore bosons and $V_{\uparrow\uparrow}'\gg V_{\downarrow\downarrow}'$ for fermions), in order to induce different intracomponent correlations $\psi\sim (z_i^{\uparrow}-z_j^{\uparrow})^{m'}(z_i^{\downarrow}-z_j^{\downarrow})^{m}$ in each component for Halperin $(m',m,n)$ states. In the following part we shall elucidate the characteristic $\mathbf{K}$ matrix from topological degeneracy, topologically invariant Chern number, fractional charge pumping, and entanglement spectrum of the ground states.

\begin{figure}[t]
  \includegraphics[height=1.75in,width=3.4in]{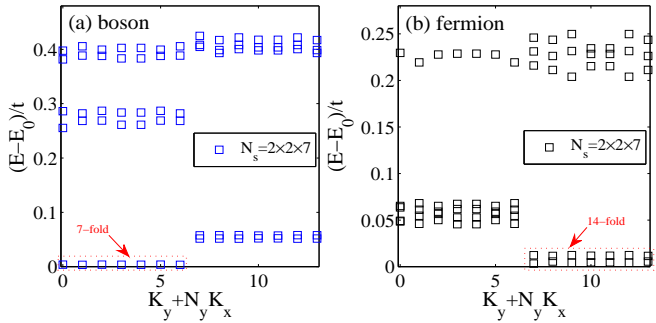}
  \caption{\label{energy} (Color online) Numerical ED results for the low energy spectrum of (a) two-component hardcore bosons at $\nu_{\uparrow}=1/7,\nu_{\downarrow}=3/7$ with $U=\infty,V_{\uparrow\uparrow}/t=100,V_{\downarrow\downarrow}=V_{\uparrow\uparrow}'=V_{\downarrow\downarrow}'=0$ and (b) two-component fermions at $\nu_{\uparrow}=1/7,\nu_{\downarrow}=2/7$ with $U=\infty,V_{\uparrow\uparrow}/t=V_{\downarrow\downarrow}/=V_{\uparrow\uparrow}'/t=100,V_{\downarrow\downarrow}'=0$ in topological checkerboard lattice.}
\end{figure}

\subsection{Ground state degeneracy}

\begin{figure}[t]
  \includegraphics[height=2.6in,width=3.4in]{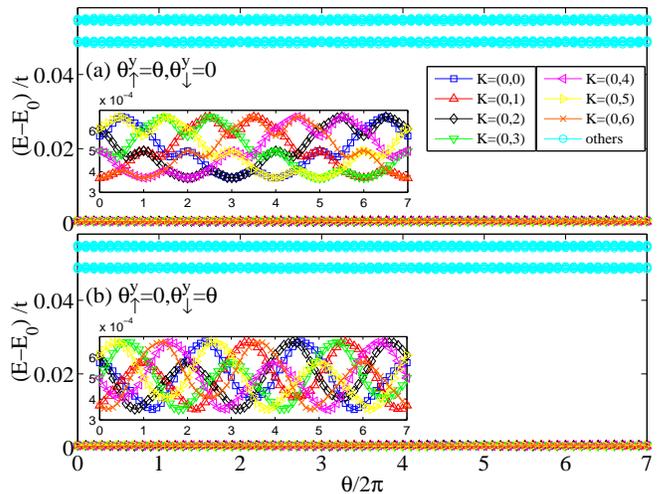}
  \caption{\label{flux421} (Color online) Numerical ED results for the low energy spectral flow of two-component hardcore bosons $N_{\uparrow}=2,N_{\downarrow}=6,N_s=2\times2\times7$ with $U=\infty,V_{\uparrow\uparrow}/t=100,V_{\downarrow\downarrow}=V_{\uparrow\uparrow}'=V_{\downarrow\downarrow}'=0$ in topological checkerboard lattice under the insertion of two types of flux quanta: (a) $\theta_{\uparrow}^{y}=\theta,\theta_{\downarrow}^{y}=0$ and (b) $\theta_{\uparrow}^{y}=0,\theta_{\downarrow}^{y}=\theta$.}
\end{figure}

First, we demonstrate the topological ground state degeneracy on finite periodic lattice using the ED study. Limited by small commensurate system sizes in ED, in order to reduce the overlap of the hopping and interaction between the next-nearest-neighbor sites on periodic torus, we choose the only available lattice geometry with commensurate $N_x=2,N_y=7$ wrapped up by the magenta quadrilateral in Fig.~\ref{lattice}(a), which is within our current ED computability. In infinite DMRG, due to the infinite cylinder length $N_x$, there is no overlap of the hopping between different sites, and the typical lattice geometry in Fig.~\ref{lattice}(b) is used.

As shown in Fig.~\ref{energy}(a) for two-component hardcore bosons in a strongly interacting regime, we find that, there exists a well-defined seven-fold quasidegenerate ground states separated from higher energy levels by a visible gap, consistent with the determinant of the $\mathbf{K}_b$ matrix. Similarly for two-component fermions in a strongly interacting regime, as indicated in Fig.~\ref{energy}(b), a low-lying manifold of fourteen-fold quasidegenerate ground states is equally distributed in seven momentum sectors.

Meanwhile we twist the boundary condition as $\psi(\rr_{\sigma}+N_{\alpha})=\psi(\rr_{\sigma})\exp(i\theta_{\sigma}^{\alpha})$ where $\theta_{\sigma}^{\alpha}$ is the twisted angle of spin-$\sigma$ particles in the $\alpha=x,y$ direction. Physically $\theta_{\sigma}^{\alpha}$ induces a momentum shift $\theta_{\sigma}^{\alpha}/N_{\alpha}$ to each particle. We further calculate the low energy spectral flow under the insertion of two types of flux quanta.
For two types of flux quanta in different components of hardcore bosons, as shown in Figs.~\ref{flux421}(a) and~\ref{flux421}(b), we find that these seven-fold ground states evolve into each other without mixing with the higher levels,
and the system returns back to itself upon the insertion of seven flux quanta for both $\theta_{\uparrow}^{\alpha}=\theta,\theta_{\downarrow}^{\alpha}=0$
and $\theta_{\uparrow}^{\alpha}=0,\theta_{\downarrow}^{\alpha}=\theta$.

\begin{figure}[t]
  \includegraphics[height=2.6in,width=3.4in]{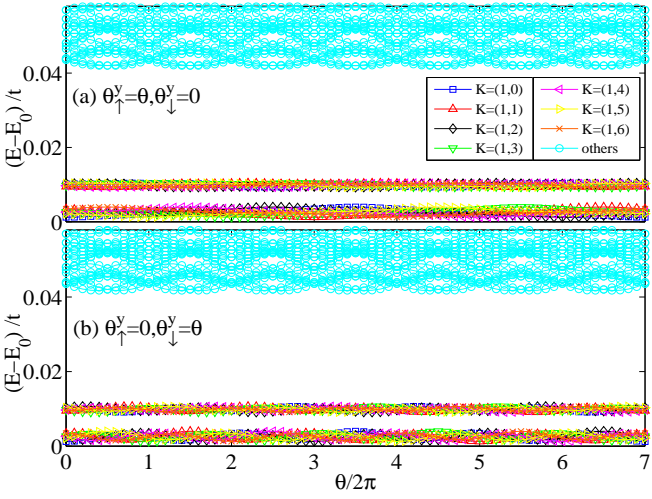}
  \caption{\label{flux531} (Color online) Numerical ED results for the low energy spectral flow of two-component fermions $N_{\uparrow}=2,N_{\downarrow}=4,N_s=2\times2\times7$ with $U=\infty,V_{\uparrow\uparrow}/t=V_{\downarrow\downarrow}/=V_{\uparrow\uparrow}'/t=100,V_{\downarrow\downarrow}'=0$ in topological checkerboard lattice under the insertion of two types of flux quanta: (a) $\theta_{\uparrow}^{y}=\theta,\theta_{\downarrow}^{y}=0$ and (b) $\theta_{\uparrow}^{y}=0,\theta_{\downarrow}^{y}=\theta$.}
\end{figure}

Similarly when applying the same twisted boundary conditions to two-component fermions, we plot the low energy spectral flow when tuning two distinct types of flux quanta in different components in Figs.~\ref{flux531}(a) and~\ref{flux531}(b). In different momentum sectors, the fourteen-fold quasidegenerate ground states evolve into each other upon flux insertion. Due to the mutual coupling between two states in the same momentum sector, they avoid level mixing.
The mutual evolution behavior is a manifestation of the topological equivalence of these ground states in each system, implying the fractionally quantized structure.

\subsection{Chern number matrix}

\begin{figure}[t]
  \includegraphics[height=2.7in,width=3.4in]{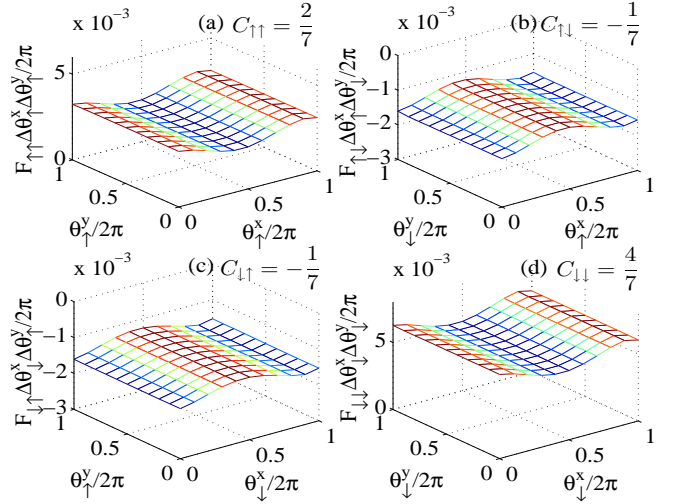}
  \caption{\label{berry421} (Color online) Numerical ED results for Berry curvatures $F_{\sigma,\sigma'}^{xy}\Delta\theta_{\sigma}^{x}\Delta\theta_{\sigma'}^{y}/2\pi$ of the ground state at momentum $K=(0,0)$ of two-component hardcore bosons $N_{\uparrow}=2,N_{\downarrow}=6,N_s=2\times2\times7$ with $U=\infty,V_{\uparrow\uparrow}/t=100,V_{\downarrow\downarrow}=V_{\uparrow\uparrow}'=V_{\downarrow\downarrow}'=0$ in topological checkerboard lattice under different twisted angles: (a) $(\theta_{\uparrow}^{x},\theta_{\uparrow}^{y})$, (b) $(\theta_{\uparrow}^{x},\theta_{\downarrow}^{y})$, (c) $(\theta_{\downarrow}^{x},\theta_{\uparrow}^{y})$, (d) $(\theta_{\downarrow}^{x},\theta_{\downarrow}^{y})$.}
\end{figure}

For any FQH state, it is common knowledge that
its Hall conductance should be a quantized value $\sigma_H=\nu$, recognized as a topological invariant (so called Chern number)~\cite{Kohmoto1985}. Hence we begin to analyze the fractionally quantized topological Chern number of the many-body ground state for interacting systems~\cite{Niu1985}. For two-component systems, we can introduce the Chern number matrix $\mathbf{C}=\begin{pmatrix}
C_{\uparrow\uparrow} & C_{\uparrow\downarrow} \\
C_{\downarrow\uparrow} & C_{\downarrow\downarrow} \\
\end{pmatrix}$ with the intercomponent response $C_{\sigma\sigma'}$ ($\sigma\neq\sigma'$) included~\cite{Sheng2003,Sheng2006}. In the parameter plane of two independent twisted angles $\theta_{\sigma}^{x}\subseteq[0,2\pi],\theta_{\sigma'}^{y}\subseteq[0,2\pi]$, we can also define the Chern number of the many-body ground state wavefunction $\psi(\theta_{\sigma}^{x},\theta_{\sigma'}^{y})$ as an integral $C_{\sigma\sigma'}=\int\int d\theta_{\sigma}^{x}d\theta_{\sigma'}^{y}F_{\sigma\sigma'}(\theta_{\sigma}^{x},\theta_{\sigma'}^{y})/2\pi$, where the Berry curvature is given by
\begin{align}
  F_{\sigma\sigma'}(\theta_{\sigma}^{x},\theta_{\sigma'}^{y})=\mathbf{Im}\left(\langle{\frac{\partial\psi}{\partial\theta_{\sigma}^x}}|{\frac{\partial\psi}{\partial\theta_{\sigma'}^y}}\rangle
-\langle{\frac{\partial\psi}{\partial\theta_{\sigma'}^y}}|{\frac{\partial\psi}{\partial\theta_{\sigma}^x}}\rangle\right).\nonumber
\end{align}

Numerically, we divide the continuous parameter plane $(\theta_{\sigma}^{x},\theta_{\sigma'}^{y})$ into $(m+1)\times(m+1)$ coarsely discretized mesh points $(\theta_{\sigma}^{x},\theta_{\sigma'}^{y})=(2k\pi/m,2l\pi/m)$ where $0\leq k,l\leq m$. We can first define the Berry connection of the wavefunction between two neighboring points as $A_{k,l}^{\pm x}=\langle\psi(k,l)|\psi(k\pm1,l)\rangle$, $A_{k,l}^{\pm y}=\langle\psi(k,l)|\psi(k,l\pm1)\rangle$.
Then the Berry curvature on the small Wilson loop plaquette $(k,l)\rightarrow(k+1,l)\rightarrow(k+1,l+1)\rightarrow(k,l+1)\rightarrow(k,l)$ is given by the gauge-invariant expression $F_{\sigma\sigma'}(\theta_{\sigma}^{x},\theta_{\sigma'}^{y})\times4\pi^2/m^2=\mathbf{Im}\ln
\big[A_{k,l}^{x}A_{k+1,l}^{y}A_{k+1,l+1}^{-x}A_{k,l+1}^{-y}\big]$.

For a given ground state at momentum $K$, by numerically calculating the Berry curvatures using $m\times m$ mesh Wilson loop plaquette in the boundary phase space with $m\geq10$, we obtain the quantized topological invariant $C_{\sigma\sigma'}$ as a summation over these discretized Berry curvatures. In our ED study of finite system sizes for two-component hardcore bosons at fillings $\nu_{\uparrow}=1/7,\nu_{\downarrow}=3/7$, as indicated in Fig.~\ref{berry421},
we obtain the smooth Berry curvature for the ground state at $K=(0,0)$ and find that it hosts the fractionally quantized Chern number matrix, namely the inverse of the $\mathbf{K}_b=\begin{pmatrix}
4 & 1 \\
1 & 2 \\
\end{pmatrix}$ matrix,
\begin{align}
  \mathbf{C}_b=\begin{pmatrix}
C_{\uparrow\uparrow} & C_{\uparrow\downarrow} \\
C_{\downarrow\uparrow} & C_{\downarrow\downarrow} \\
\end{pmatrix}=\frac{1}{7}\begin{pmatrix}
2 & -1 \\
-1 & 4 \\
\end{pmatrix}=\mathbf{K}_b^{-1}.\label{chernb}
\end{align}

\begin{figure}[t]
  \includegraphics[height=2.6in,width=3.4in]{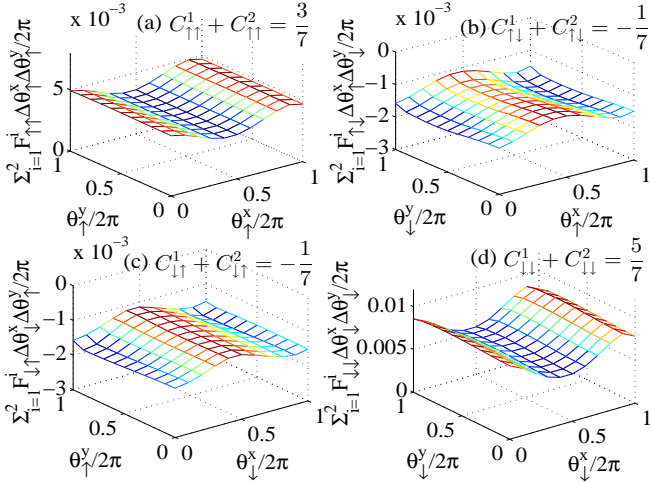}
  \caption{\label{berry531} (Color online) Numerical ED results for Berry curvatures $F_{\sigma,\sigma'}^{xy}\Delta\theta_{\sigma}^{x}\Delta\theta_{\sigma'}^{y}/2\pi$ of the two-fold quasidegenerate ground states at momentum $K=(\pi,0)$ of two-component fermions $N_{\uparrow}=2,N_{\downarrow}=4,N_s=2\times2\times7$ with $U=\infty,V_{\uparrow\uparrow}/t=V_{\downarrow\downarrow}/=V_{\uparrow\uparrow}'/t=100,V_{\downarrow\downarrow}'=0$ in topological checkerboard lattice under different twisted angles: (a) $(\theta_{\uparrow}^{x},\theta_{\uparrow}^{y})$, (b) $(\theta_{\uparrow}^{x},\theta_{\downarrow}^{y})$, (c) $(\theta_{\downarrow}^{x},\theta_{\uparrow}^{y})$, (d) $(\theta_{\downarrow}^{x},\theta_{\downarrow}^{y})$.}
\end{figure}

Similarly, for two-component fermions at fillings $\nu_{\uparrow}=1/7,\nu_{\downarrow}=2/7$, from the ED study of finite system sizes, we find that for the two coupled ground states at momentum $K=(\pi,0)$, the total Berry curvature $\sum_{i=1}^2F_{\sigma\sigma'}^i$ is also smooth and they share a total quantized Chern number $\sum_{i=1}^2C_{\sigma\sigma'}^i$, as indicated in Fig.~\ref{berry531}. For each of these ground states, it hosts the average many-body Chern number matrix
\begin{align}
  \mathbf{C}_f=\begin{pmatrix}
C_{\uparrow\uparrow} & C_{\uparrow\downarrow} \\
C_{\downarrow\uparrow} & C_{\downarrow\downarrow} \\
\end{pmatrix}=\frac{1}{14}\begin{pmatrix}
3 & -1 \\
-1 & 5 \\
\end{pmatrix}=\mathbf{K}_f^{-1}\label{chernf}
\end{align}
in correspondence to the inverse of the $\mathbf{K}_f=\begin{pmatrix}
5 & 1 \\
1 & 3 \\
\end{pmatrix}$ matrix.

\subsection{DMRG study}\label{dmrg}

Following the last section, we move on to discuss the topological nature from the perspective of DMRG approach, focusing on bosonic Halperin $(4,2,1)$ FQH effect. Here we exploit an unbiased DMRG simulation of charge pumping, chiral edge mode and topological entanglement entropy on an infinite cylindrical geometry with different widths.

\textit{Charge pumping}---According to Laughlin's arguments~\cite{Laughlin1981}, quantized charge pumping via an adiabatically periodic perturbation is a hallmark of the two-dimensional quantum Hall effect, which is connected to the Hall conductance. In Eqs.~\ref{chernb} and~\ref{chernf}, the remarkable feature is that two-component quantum Hall system can host a quantized drag Hall conductance, quantitatively determined by the symmetric matrix elements $C_{\uparrow\downarrow}=C_{\downarrow\uparrow}$. To simulate this physical effect, we can utilize DMRG to calculate the charge pumping on infinite cylinder systems as the twisted angle $\theta_{\sigma}^{\alpha}$ changes from zero to $2\pi$~\cite{Gong2014}, which leads to a change of lattice momentum $k_{\alpha}\rightarrow k_{\alpha}+2\pi/N_{\alpha}$ in the $\sigma$-component particles. If one flux quantum $\Delta\theta_{\sigma}^{\alpha}=2\pi$ is inserted (equivalent to the momentum change $\Delta k_{\alpha}=2\pi/N_{\alpha}$), the quantized charge pumping is obtained, equivalent to the charge Hall conductance.

\begin{figure}[t]
  \includegraphics[height=2.0in,width=3.4in]{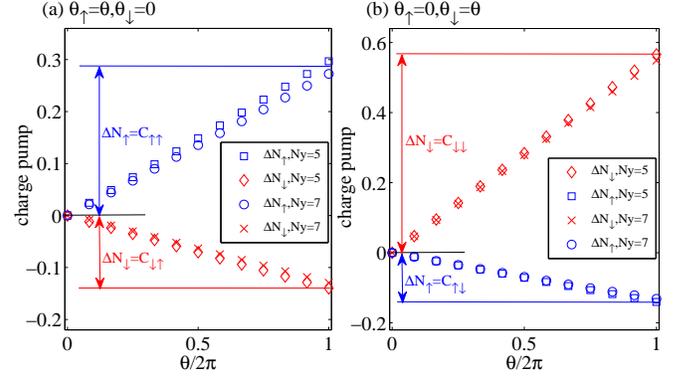}
  \caption{\label{pump} (Color online) Fractional charge transfers for two-component hardcore bosons at $\nu_{\uparrow}=1/7,\nu_{\downarrow}=3/7$ with $U=\infty,V_{\uparrow\uparrow}/t=50,V_{\downarrow\downarrow}=V_{\uparrow\uparrow}'=V_{\downarrow\downarrow}'=0$ on the cylinder of topological checkerboard lattice under two types of the insertions of flux quantum: (a) $\theta_{\uparrow}^{y}=\theta,\theta_{\downarrow}^{y}=0$ and (b) $\theta_{\uparrow}^{y}=0,\theta_{\downarrow}^{y}=\theta$. Here infinite DMRG is used with different cylinder widths $N_y=5,7$, keeping the maximal bond dimension up to 10000.}
\end{figure}

Numerically we cut the cylinder into left-half and right-half parts along the $x$ direction, $|\psi\rangle=\sum_{a,b}\psi_{ab}|a\rangle_L\bigotimes|b\rangle_R$ where $|a\rangle_L,|b\rangle_R$ are the corresponding basis states of the left-half and right-half parts. By tracing over the right-half part of the system, we obtain the reduced density matrix $\widehat{\rho}_L=\sum_b\langle b|\psi\rangle\langle\psi|b\rangle=\sum_{a,a'}\rho_{aa'}|a\rangle\langle a'|$ in the basis state $|a\rangle_L$ of the left-half part with matrix elements $\rho_{aa'}=\sum_b\psi_{a'b}^{\ast}\psi_{ab}$, classified by the charge quantum numbers $\Delta Q_{\uparrow}=N_{\uparrow}^a-N_{\uparrow}/2,\Delta Q_{\downarrow}=N_{\downarrow}^a-N_{\downarrow}/2$ (these values are measured as the deviation of the particle number $N_{\sigma}^a$ of $|a\rangle_L$ from the equal partition $N_{\sigma}^a=N_{\sigma}^b=N_{\sigma}/2$). When the flux quantum $\theta_{\sigma'}^{y}$ of the $\sigma'$-component particle is inserted along the $y$ direction, the net charge transfers for the $\sigma$-component particle from the right-half side to the left-half side on the cylinder are encoded by the average value $N_{\sigma}(\theta_{\sigma'}^{y})=tr[\widehat{\rho}_L(\theta_{\sigma'}^{y})\widehat{N}_{\sigma}]$
as a function of $\theta_{\sigma'}^{y}$. Previously it have been shown~\cite{Zeng2017,Zeng2018} that for multicomponent FQH effect, we can obtain a well-defined drag charge response between different components by inserting one flux quantum $\theta_{\sigma}^{y}=\theta,\theta_{\sigma'}^{y}=0$ in only one component from $\theta=0$ to $\theta=2\pi$, which even holds for Bose-Fermi mixtures~\cite{Zeng2021}. As illustrated in Figs.~\ref{pump}(a) and~\ref{pump}(b), we use two different types of flux quanta adopted in Fig.~\ref{flux421}, and find that (i) for the flux thread $\theta_{\uparrow}=\theta,\theta_{\downarrow}=0,\theta\subseteq[0,2\pi]$, the charge pumpings
\begin{align}
  &\Delta N_{\uparrow}=N_{\uparrow}(2\pi)-N_{\uparrow}(0)\simeq C_{\uparrow\uparrow}=\frac{2}{7}, \nonumber\\
  &\Delta N_{\downarrow}=N_{\downarrow}(2\pi)-N_{\downarrow}(0)\simeq C_{\downarrow\uparrow}=-\frac{1}{7} \nonumber
\end{align}
and (ii) for the flux thread $\theta_{\uparrow}=0,\theta_{\downarrow}=\theta,\theta\subseteq[0,2\pi]$, the charge pumpings
\begin{align}
  &\Delta N_{\uparrow}=N_{\uparrow}(2\pi)-N_{\uparrow}(0)\simeq C_{\uparrow\downarrow}=-\frac{1}{7},\nonumber\\
  &\Delta N_{\downarrow}=N_{\downarrow}(2\pi)-N_{\downarrow}(0)\simeq C_{\downarrow\downarrow}=\frac{4}{7} \nonumber
\end{align}
in excellent agreement with the expectations of the Chern number matrix in Eq.~\ref{chernb} for two-component hardcore bosons at fillings $\nu_{\uparrow}=1/7,\nu_{\downarrow}=2/7$.

\begin{table}[t]
\setlength{\tabcolsep}{0.2cm}
\caption{\label{counting} The energy levels of edge mode.}
\begin{tabular}{c|c|c}
\hline
\hline
Levels & mode & ${n_{m}^{c},n_{m}^{s}}$ \\
\hline
$\Delta K=0$ & - & $n_{m}^{c}=0,n_{m}^{s}=0$ \\
\hline
$\Delta K=1$ & charge & $n_{1}^{c}=1,n_{m\neq1}^{c}=0,n_{m}^{s}=0$ \\
$\Delta K=1$ & spin & $n_{m}^{c}=0,n_{1}^{s}=1,n_{m\neq1}^{s}=0$ \\
\hline
$\Delta K=2$ & charge & $n_{2}^{c}=1,n_{m\neq2}^{c}=0,n_{m}^{s}=0$ \\
$\Delta K=2$ & charge & $n_{1}^{c}=2,n_{m\neq1}^{c}=0,n_{m}^{s}=0$ \\
$\Delta K=2$ & mixed & $n_{1}^{c}=1,n_{m\neq1}^{c}=0,n_{1}^{s}=1,n_{m\neq1}^{s}=0$ \\
$\Delta K=2$ & spin & $n_{m}^{c}=0,n_{1}^{s}=2,n_{m\neq1}^{s}=0$ \\
$\Delta K=2$ & spin & $n_{m}^{c}=0,n_{2}^{s}=1,n_{m\neq2}^{s}=0$ \\
\hline
\end{tabular}
\end{table}

\textit{Chiral edge modes}---Further, we calculate the low-lying momentum-resolved entanglement spectrum of an infinite cylinder, defined as $\xi_i=-\ln\rho_i$ where $\rho_i$ is the eigenvalue of the density matrix $\widehat{\rho}_L$, which would reveal the characteristic chirality and excited level counting of edge modes~\cite{Li2008}. For edge modes of Abelian two-component FQH effect with the $\mathbf{K}=\begin{pmatrix}
m' & n\\
n & m\\
\end{pmatrix}$ matrix, they are described by free bosonic operator (namely collective density or spin operators) which conserved the charge and spin quantum numbers, and the edge chirality is determined by the signs of the eigenvalues $\left[m'+m\pm\sqrt{(m'-m)^2+4n^2}\right]/2$ of the $\mathbf{K}$ matrix~\cite{Wen1992edge,Wen1995}. For $\mathbf{K}=\begin{pmatrix}
4 & 1\\
1 & 2\\
\end{pmatrix}$, there are two parallel chiral edge modes. In field description~\cite{Furukawa2013}, for given charge sectors $\Delta Q_{\uparrow},\Delta Q_{\downarrow}$, apart from certain constant, these two edge modes are approximately described by the edge Hamiltonian $H_{edge}=2\pi/N_y\times(v_cK^c+v_sK^s)$, and the corresponding momentum operator $K=K_0+2\pi/N_y\times(K^c+K^s)$ with $K^{c(s)}=\sum_{m=1}^{\infty}mn_m^{c(s)}$ where ${n_{m}^{c(s)}}$ denotes the set of non-negative integers of edge charge (spin) excitation mode. We can simulate the excitation level in momentum space, and derive the degenerate level pattern of ${n_{m}^{c(s)}}$ for any momentum $\Delta K=K-K_0=\sum_{m=1}^{\infty}m(n_m^c+n_m^s)$ in unit of $2\pi/N_y$. For example, as indicated in Table~\ref{counting}, the zeroth excited level corresponds to $\Delta K=0$ with only one possibility; the first excited level corresponds to $\Delta K=1$ with two possibilities; the second excited level corresponds to $\Delta K=2$ with five possibilities.

\begin{figure}[t]
  \includegraphics[height=2.12in,width=3.4in]{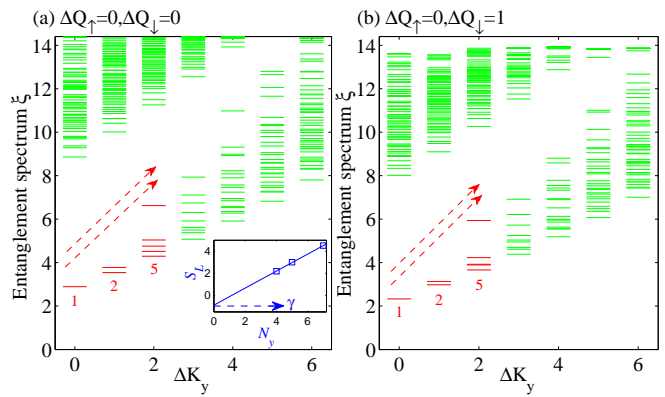}
  \caption{\label{es} (Color online) Chiral edge mode identified from the momentum-resolved entanglement spectrum for two-component hardcore bosons at $\nu_{\uparrow}=1/7,\nu_{\downarrow}=3/7$ with $U=\infty,V_{\uparrow\uparrow}/t=50,V_{\downarrow\downarrow}=V_{\uparrow\uparrow}'=V_{\downarrow\downarrow}'=0$ on the cylinder of topological checkerboard lattice. The horizontal axis shows the relative momentum $\Delta K_y=K_y-K_{y}^{0}$ (in units of $2\pi/N_y$). Numbers below the dashed red line label the excitation level counting $1,2,5$ at different momenta for typical charge sectors (a) $\Delta Q_{\uparrow}=0,\Delta Q_{\downarrow}=0$ and (b) $\Delta Q_{\uparrow}=0,\Delta Q_{\downarrow}=1$. The small inset depicts the infinite DMRG extraction of topological entanglement entropy $\gamma$ using the area law $S_L(N_y)=2\alpha N_y-\gamma$. Here the maximal bond dimension is kept up to $10000$. }
\end{figure}

In Figs.~\ref{es}(a) and~\ref{es}(b), we plot the structure of the momentum-resolved entanglement spectrum on the infinite cylinder with width $N_y=7$ for different charge sectors. We observe two forward-moving branches of the low-lying bulk entanglement spectrum with charge and spin branches mixed, and obtain the low-lying level counting $1,2,5,\cdots$ separated by a large entanglement gap from the higher levels, matching with the gapless nature of two gapless free bosonic edge theories.

\textit{Topological entanglement entropy}---Meanwhile, for three different cylinder widths $N_y=4,5,7$, we also calculate the entanglement entropy of the left-half part as $S_L(\ell)=-\sum_i\rho_i\ln\rho_i$ with the boundary length $\ell=2N_y$. We extract the topological entanglement entropy $\gamma$ via the area law of half-cylinder entanglement entropy $S_L(\ell)=\alpha \ell-\gamma$. The topological entanglement entropy $\gamma=\ln(D)$ is a universal constant with $D$ the total quantum dimension of Abelian anyons~\cite{Kitaev2006,Levin2006}, i.e. $D=\sqrt{m'm-n^2}$ for Halperin $(m',m,n)$ FQH liquid. The linear fit in the inset of Fig.~\ref{es}(a) indeed give a topological entanglement entropy $\gamma\sim0.89$ which is remarkably close to the theoretical value $\ln\sqrt{7}$ of bosonic Halperin $(4,2,1)$ FQH effect.

\section{Conclusion}\label{summary}

To summarize, we have introduced a microscopic topological $\pi$-flux checkerboard lattice model of strongly interacting two-component quantum particles (bosons and fermions), and numerically demonstrated the emergence of bosonic Halperin $(4,2,1)$ FQH effect with the $\mathbf{K}=\begin{pmatrix}
4 & 1\\
1 & 2\\
\end{pmatrix}$ matrix at fillings $\nu_{\uparrow}=1/7,\nu_{\downarrow}=3/7$, and fermionic Halperin $(5,3,1)$ FQH effect with the $\mathbf{K}=\begin{pmatrix}
5 & 1\\
1 & 3\\
\end{pmatrix}$ matrix at fillings $\nu_{\uparrow}=1/7,\nu_{\downarrow}=2/7$ through the ED study. We find two pieces of major evidence: (i) the $\det|\mathbf{K}|$-fold ground-state degeneracies equivalent to the determinant of the $\mathbf{K}$ matrix, (ii) fractionally quantized topological Chern number matrix equivalent to the inverse of the $\mathbf{K}$ matrix. Our comprehensive DMRG studies of fractional charge pumping, two parallel-propagating chiral edge branches and topological entanglement entropy further confirm two-component topological nature of bosonic Halperin $(4,2,1)$ FQH effect at fillings $\nu_{\uparrow}=1/7,\nu_{\downarrow}=3/7$. We remark that our flat band models of multicomponent systems with strong intercomponent and intracomponent interactions might furthermore serve as a promising paradigm for engineering exotic quantum Hall physics with an imbalanced spin polarization.

\begin{acknowledgements}
T.S.Z thanks D. N. Sheng and W. Zhu for inspiring discussions and prior collaborations on multicomponent fractional quantum Hall physics in topological flat band models.
This work is supported by the National Natural Science Foundation of China (NSFC) under Grant No. 12074320.
\end{acknowledgements}

\end{document}